\documentclass[a4paper]{jpconf}
\usepackage[utf8]{inputenc}
\usepackage{graphicx}
\usepackage{caption}
\usepackage{subfigure}
\usepackage{amsfonts,amsmath,amssymb}
\usepackage{multirow}
\usepackage{braket}
\usepackage{float}
\usepackage{epstopdf}
\usepackage{tikz}
\usepackage[numbers,square,sort&compress]{natbib}
\usetikzlibrary{arrows,patterns,decorations.pathreplacing,shapes.geometric}
\usepackage{enumitem}
\usepackage{dcolumn}

\newcolumntype{d}[1]{D{.}{.}{#1}}
\begin{document}
\title{Towards efficient density functional theory calculations without self-interaction: The Fermi-L\"owdin orbital self-interaction correction}
\author{K. A. Jackson, J. E. Peralta, R. P. Joshi, K. P. Withanage, K. Trepte, K. Sharkas and A. I. Johnson}
\address{Physics Department and Science of Advanced Materials Ph.D. Program, Central Michigan University, Mt. Pleasant, MI USA}
\ead{jacks1ka@cmich.edu, peral1j@cmich.edu}
\date{October 8, 2018}
\begin{abstract}
    The Fermi-L\"owdin orbital (FLO) approach to the Perdew-Zunger self-interaction correction (PZ-SIC) to density functional theory (DFT) is described and an improved approach to the problem of optimizing the Fermi-orbitals in order to minimize the DFT-SIC total energy is introduced.  To illustrate the use of the FLO-SIC method, results are given for several applications involving problems where self-interaction errors are pronounced.
\end{abstract}


\section{Introduction}
Density functional theory (DFT) is the workhorse of computational condensed matter physics.  Nowhere is this more evident than in a materials genome project where DFT calculations, 
typically using the Perdew-Burke-Ernzerhof (PBE)~\cite{perdew1996generalized} generalized gradient approximation (GGA) to the exchange-correlation functional, are performed for 
thousands of materials to create 
a large database of calculated results that can be mined to discover property trends.  A number of these projects were discussed 
at CCP2018. PBE is used for these calculations because it combines reliability with efficiency, \textit{i.e.}  accurate predictions of materials properties can be produced at relatively low computational cost.  A subtle factor in the success of this enterprise is that the calculations are done for atomic configurations at or near equilibrium.  In such situations PBE and similar functionals generally perform well; however, the reliability of these methods deteriorates when the atoms are far from equilibrium. For example, chemical transition states typically involve strongly stretched bonds and PBE and other functionals significantly underestimate the corresponding reaction barriers.  The culprit behind this breakdown is self-interaction error (SIE), which is present in all approximate semilocal density functionals due to an inexact cancellation of self-Coulomb and self-exchange-correlation energies.  In addition to too-low reaction barriers, SIE causes a raft of other problems in DFT calculations, from orbital energies that are too high, to problems in the description of charge transfer.  
\newline

The Perdew-Zunger self-interaction correction (PZ-SIC)~\cite{PZ} was introduced in the 1980's to remove electron self-interaction from DFT.  Early tests indicated that many SIE-related problems were indeed fixed by PZ-SIC, but the method was not widely adopted for two reasons.  First, the method is computationally demanding relative to uncorrected DFT, and second, while the method can improve the success of DFT in situations where SIE is important, it degrades the performance of PBE and related functionals in settings where DFT already works well~\cite{vydrov,G2}. 
\newline

The Fermi-L\"{o}wdin orbital self-interaction correction (FLO-SIC)~\cite{pederson2014communication,pederson2015fermi,full_selfconsistent} is a recent implementation of PZ-SIC that offers a potential computational advantage over earlier approaches.  A long-term goal of our research is to use FLO-SIC to explore new approaches of combining SIC with DFT to remove the effects of SIE without diminishing the otherwise good performance of DFT. A first step in this direction is to more fully assess the method for problems where SIE is known to be important.  In this paper we present a brief overview of the FLO-SIC formalism, including a discussion of recent methodological developments that improve its efficiency and functionality.  We then present illustrative examples to demonstrate how FLO-SIC successfully addresses the effects of electron self-interaction.  We conclude with a reflection on remaining challenges to implementing FLO-SIC more broadly.

\section{Theoretical background}




In DFT, the electrostatic Coulomb energy of the electrons $E_{\text{H}}$ includes the interaction of each electron with itself.  
In exact DFT, this self-Coulomb energy is exactly cancelled by the exchange-correlation (XC) energy $E_{\text{XC}}$, but residual self-interaction remains when an 
approximate XC functional is used instead. In PZ-SIC, this residual energy is subtracted out on an orbital-by-orbital basis~\cite{PZ}
\begin{equation}
\label{eq:PZ}
    E^{\text{DFT-SIC}} = E^{\text{DFT}}[n^{\alpha},n^{\beta}] - \sum_{i\sigma}(E_{\text{XC}}[n_{i}^{\sigma},0] + E_{\text{H}}[n_{i}^{\sigma}]) \, ,
\end{equation}
where $n_{i}^{\sigma}$ are single orbital densities of spin $\sigma$, $E^{\text{DFT}}$ is the DFT energy functional, $E_{\text{XC}}$ is the exchange-correlation functional used and 
$E_{\text{H}}$ is the Hartree, or Coulomb, energy
\begin{equation}
   E_{\text{H}} = \frac{1}{2} \int \frac{n(\textbf{r}) n(\textbf{r}')}{|\textbf{r} - \textbf{r}'|} \text{d}^{3} r \text{d}^{3} r'.
\end{equation}

Schr\"{o}dinger-like DFT-SIC equations are derived analogously to the Kohn-Sham (KS) equations using a standard variational method, yielding
\begin{equation}
 \left(\hat{H}_{\sigma}^{\text{DFT}} + V_{i\sigma}^{\text{SIC}}\right) \phi_{i\sigma} = 
 \hat{H}_{i\sigma}^{\text{DFT-SIC}} \phi_{i\sigma} = \sum_{j = 1}^{M_{\sigma}} \lambda_{ji}^{\sigma} \phi_{j\sigma}.
 \label{scf}
\end{equation}
Here $\hat{H}_{\sigma}^{\text{DFT}}$ is the DFT Hamiltonian, $V_{i\sigma}^{\text{SIC}}$ is the SIC potential of orbital~$i$, and $\hat{H}_{i}^{\text{DFT-SIC}}$ is the corresponding 
SIC Hamiltonian. The $\lambda_{ji}^{\sigma}$ are the Lagrange multipliers introduced in the variational procedure to ensure orthogonality between the $M_{\sigma}$ occupied spin orbitals.  
\newline

A challenging feature of PZ-SIC is that the total energy $E^{\text{DFT-SIC}}$ and the DFT-SIC equations are orbital-dependent.  This means that individual orbital densities must be optimized in order to obtain the lowest $E^{\text{DFT-SIC}}$.  The key idea behind FLO-SIC is to use localized Fermi orbitals (FO) in the PZ energy functional~\cite{pederson2014communication,pederson2015fermi,full_selfconsistent}. 
A FO is defined as
\begin{equation}
 F_{i \sigma} = \sum_{j=1}^{M_{\sigma}}\frac{\psi_{j \sigma}^{*}(\mathbf{a}_{i \sigma})\psi_{j \sigma}(\textbf{r})}{\sqrt{\sum_{j=1}^{M_{\sigma}} |\psi_{j \sigma}(\mathbf{a}_{i \sigma})|^2}}\, ,
\end{equation}
where $\psi_{j \sigma}(\textbf{r})$ is any set of orbitals that spans the occupied orbital space (\textit{i.e.} $\sum_{j=1}^{M_{\sigma}} |\psi_{j \sigma}(\mathbf{r})|^2 = n_{\sigma}(\textbf{r})$).
The ${\textbf{a}}_{i\sigma}$ are called Fermi-orbital descriptors (FODs). They are $M_{\sigma}$ distinct points in space that characterize the transformation.  The FO are normalized, but not orthogonal, so the Löwdin method~\cite{loewdin1950} is applied to obtain orthonormal Fermi-L\"{o}wdin orbitals $\phi_{i\sigma}$ (FLOs). 
\newline

\begin{figure}[h] %
 \centering
 \begin{tikzpicture} [scale=1.2,>=triangle 45]
 \draw[-,very thick] (-2.2,+0.6) rectangle (2.0,+0.1);
 \node at (0.0,+0.35) {\bf{Initial FOD positions} \textbf{a}$_{i}$};
 \draw[->,very thick] (0.0,0.10) -- (0.0,-0.65);
 
 \draw[-,very thick] (-2.9,-0.65) rectangle (2.9,-1.15);
 \node at (0.0,-0.90) {\bf{Generate Fermi-L\"owdin orbitals $\ket{\phi_{i}}$}};
 \draw[->,very thick] (0.0,-1.15) -- (0.0,-1.90);
 
 \draw[-,very thick] (-2.0,-1.90) rectangle (2.0,-2.50);
 \node at (0.0,-2.20) {\bf{Calculate \{$\hat{H}_{i}^{\text{DFT-SIC}}$\}}};
 \draw[->,very thick] (0.0,-2.50) -- (0.0,-3.25);

 \draw[-,very thick] (-2.4,-3.25) rectangle (2.4,-3.95);
 \node at (0.0,-3.60) {\bf{Satisfy $\hat{H}_{i}\ket{\phi_{i}} = \sum_{j = 1}^{N} \lambda_{ij}\ket{\phi_{j}}$}};
 \draw[->,very thick] (0.0,-3.95) -- (0.0,-4.70);
 
 \node [draw, very thick, diamond, aspect=4] at (0.0,-5.35) {\bf{Density converged?}};
 \draw[-,very thick]  (-2.5,-5.35)  --node[above=-1pt] {\bf{no}} (-4.0,-5.35);
 \draw[-,very thick]  (-3.98,-5.35) -- (-3.98,-0.90);
 \draw[->,very thick] (-4.00,-0.90) -- (-2.90,-0.90);
 \draw[->,very thick] (0.0,-6.00) --node[right=1pt] {\bf{yes}} (0.0,-6.75);
 
 \draw[-,very thick] (-2.4,-6.75) rectangle (2.4,-7.25);
 \node at (0.0,-7.00) {\bf{Calculate FOD forces $\partial E/\partial \textbf{a}_{i}$}};
 \draw[->,very thick] (0.0,-7.25) -- (0.0,-8.00);
 
 \node [draw, very thick, diamond, aspect=4] at (0.0,-8.75) {\bf{FOD forces converged?}};
 \draw[-,very thick]  (+2.90,-8.75)  --node[above=-1pt] {\bf{no}} (+4.30,-8.75);
 \draw[-,very thick]  (+4.28,-8.75)  -- (+4.28,-5.05);
 \draw[-,very thick]  (+3.30,-5.05) rectangle (+5.30,-4.55);
 \node at (4.3,-4.80) {\bf{Update} \textbf{a}$_{i}$};
 \draw[-,very thick]  (+4.28,-4.55)  -- (+4.28,-0.90);
 \draw[->,very thick]  (+4.30,-0.90)  -- (+2.90,-0.90);
 \draw[->,very thick] (0.0,-9.50) --node[right=1pt] {\bf{yes}} (0.0,-10.25);
 
 \draw[-,very thick] (-1.0,-10.25) rectangle (1.0,-10.75);
 \node at (0.0,-10.50) {\bf{Output}}; 

 \end{tikzpicture}
\caption{General outline of the FLO-SIC method.}
\label{FLO-SIC_fig}
\end{figure}
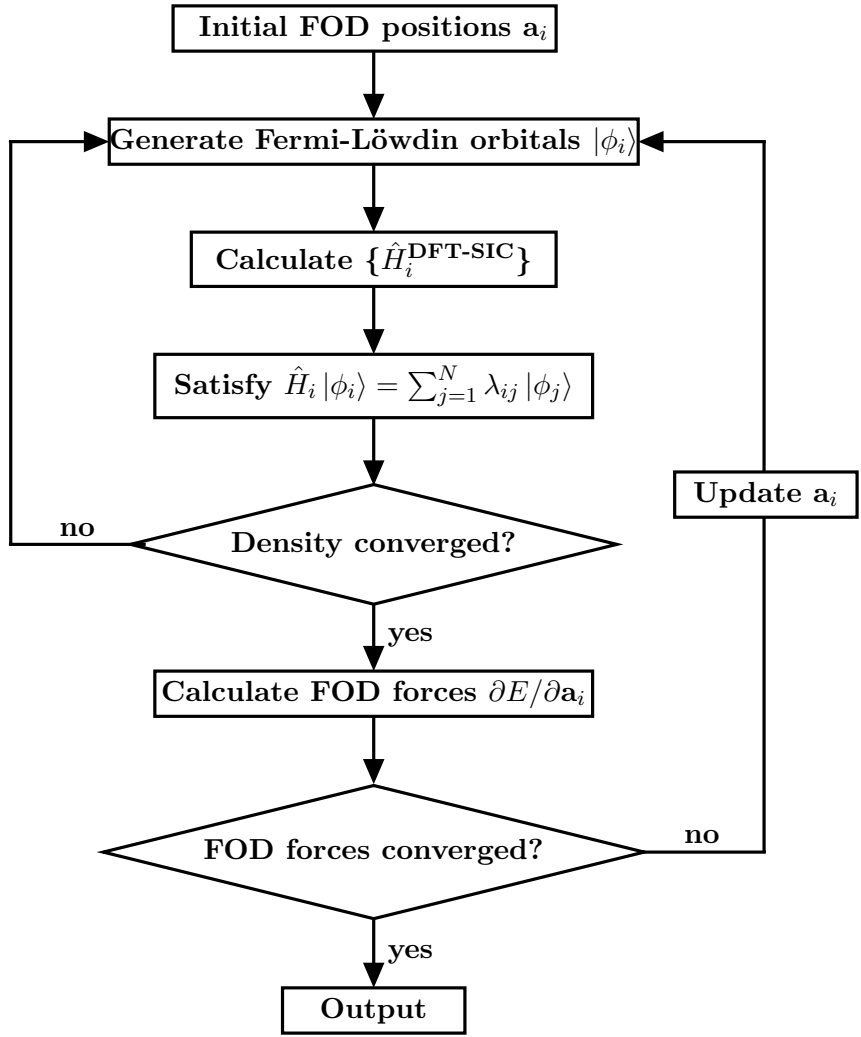

Minimizing $E^{\text{DFT-SIC}}$ requires that both the total density and the FLOs are optimized.  The latter is accomplished by rearranging the FOD positions. As discussed further below, the FOD positions can be optimized by using the gradients of the total energy with respect to the ${\textbf{a}_i}$, which we refer to as FOD forces.  A flowchart summarizing a FLO-SIC energy minimization is given in Fig.~\ref{FLO-SIC_fig}.
\newline


\section{Improving FOD optimization}
In FLO-SIC, the total energy is minimized by optimizing the FOD positions. Pederson~\cite{pederson2015fermi} derived an expression for the gradients $\textbf{g}_{i}=\partial E^{\text{DFT-SIC}}/\partial \textbf{a}_{i}$ and used these in a conjugate gradient algorithm to optimize the FODs for some closed-shell atoms. The method is akin to that used to optimize the arrangements of atoms in a molecule.  Using a similar approach, we 
determined minimum-energy FODs for all the atoms Li to Kr~\cite{atomicFODs2017}. In practice, the conjugate gradient algorithm may take hundreds of steps to determine optimal FOD positions, and the number of steps required grows as the number of FODs increases.  It is clearly important to make the FOD optimization as efficient as possible. 
\newline

To gain insight into the nature of the FOD energy surface, we computed second derivatives of the energy with respect to the FOD positions at the minimum energy FOD arrangement for several closed-shell atoms.  For a particular FOD $\textbf{a}$, the second derivative $h_{kl}= \partial^2 E^{\text{DFT-SIC}}/ \partial a_{k}\partial a_{l}$ was calculated numerically via finite differences of the corresponding gradients 

\begin{equation}
    h_{kl}=\frac{(g_k|_{+\Delta x_l}-g_k|_{-\Delta x_l})}{2\Delta x_l},
    \label{fdif}
\end{equation}

where $g_k$ is the gradient in the $k$ direction ($k = x$, $y$ or $z$) and $\pm \Delta x_{l}$ is an FOD displacement in the $l$ direction. We used displacements of $0.001$~Bohr to compute the derivatives. The average of the diagonal element ($h_D = \frac{1}{3}\sum_{k} h_{kk}$) is a measure of how steeply curved the energy surface is for the given FOD.  For a typical atom, the curvature is  similar for all FODs corresponding to local orbitals in the same electronic shell.  The averaged, per-shell values of $h_D$ for the closed-shell atoms Ne, Mg, Ar, Ca, Zn, Kr are presented in table~\ref{hessians}.

\begin{table}[ht]
\centering
\caption{Averaged diagonal elements $h_D$ of the second derivative matrix for each shell (a.u).}
\label{hessians}
\begin{tabular}{|ll|d{2.2} d{2.3} d{1.3} r|}
\hline
   & $Z$  & \multicolumn{1}{c}{1s}  & \multicolumn{1}{c}{2s2p}  & \multicolumn{1}{c}{3s3p3d}    & \multicolumn{1}{c|}{4s4p}         \\\hline
Ne & 10   &  0.52                 & 0.011                    & \multicolumn{1}{c}{---}       & \multicolumn{1}{c|}{---}          \\
Mg & 12   &  1.3                 & 0.044                    & \multicolumn{1}{c}{---}       & \multicolumn{1}{c|}{---}          \\
Ar & 18   &  6.2                 & 0.36                    & 0.012                        & \multicolumn{1}{c|}{---}          \\
Ca & 20   & 12.                & 0.92                    & 0.026                        & \multicolumn{1}{c|}{---}          \\
Zn & 30   & 42.                & 7.8                    & 0.30                        & \multicolumn{1}{c|}{---}          \\
Kr & 36   & 81.                & 18.                   & 0.81                        & 0.0022                            \\\hline
\end{tabular}
\end{table}

$h_D$ for the 1s FOD is at least an order of magnitude larger than those 
for valence FODs for all atoms. For example, for Kr the 1s second derivative (81 ~a.u.) is four orders of magnitude larger than that of the 4s4p FODs (0.0022~a.u.). This implies that for similar displacements of the FODs from equilibrium, the 1s FOD gradient would be vastly larger than a 4s4p gradient. This is a signature of an ill-conditioned optimization problem that can result in slow convergence to a solution using a straightforward application of gradient-based algorithms.  
\newline

To accelerate the convergence, we used the averaged diagonal elements $h_D$ of the second derivative matrix to scale the gradients (\textbf{g}) and the FOD coordinates (\textbf{a}) as follows
\begin{align}
    \textbf{g}^\prime & =h_D^{-1/2}\textbf{g} \\
    \textbf{a}^\prime & =h_D^{1/2}\textbf{a} ,
\end{align}

prior to using the coordinates and gradients in the optimization algorithm.  
The effect is to make the scaled second derivatives $\partial g_k^{'}/ \partial a_l^{'}$ approximately the same size for all FODs.  To carry out this preconditioning step for atoms not listed in table~\ref{hessians}, we fit the values of $h_D$ for each shell vs atomic number $Z$.  For each shell, a cubic polynomial gives an excellent fit to the data in the table.  Values for any atom can be extracted from these fits.
\newline



The scaled gradients and coordinates can be used in conjunction with any gradient-based optimization algorithm.  We used them with the conjugate gradient (CG) and L-BFGS algorithms. To compare the performance of these methods with and without  preconditioning, we considered the Kr and Cr atoms.  We used the same starting set of FODs for all methods. To create the starting sets, we displaced each FOD from its ideal position given in Ref.~[\citenum{atomicFODs2017}], while  maintaining the same approximate distance to the nucleus.  For the scaled cases, we simply applied the scaling transformations to the gradient and coordinate vectors immediately before the call to the optimization subroutine and reversed the scaling for the updated coordinate vector returned from the subroutine.  Each FOD was then moved using the unscaled update.
\newline

\begin{figure}[h]
    \centering
    \subfigure[Kr]{\includegraphics[trim={1.5cm 0.2cm 2.5cm 2.0cm},clip,scale=0.35]{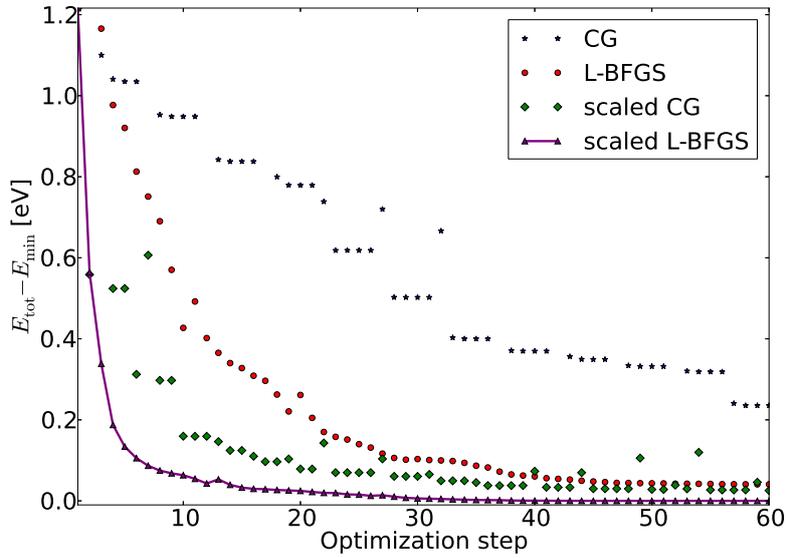}}
    \subfigure[Cr]{\includegraphics[trim={1.5cm 0.2cm 2.5cm 2.0cm},clip,scale=0.35]{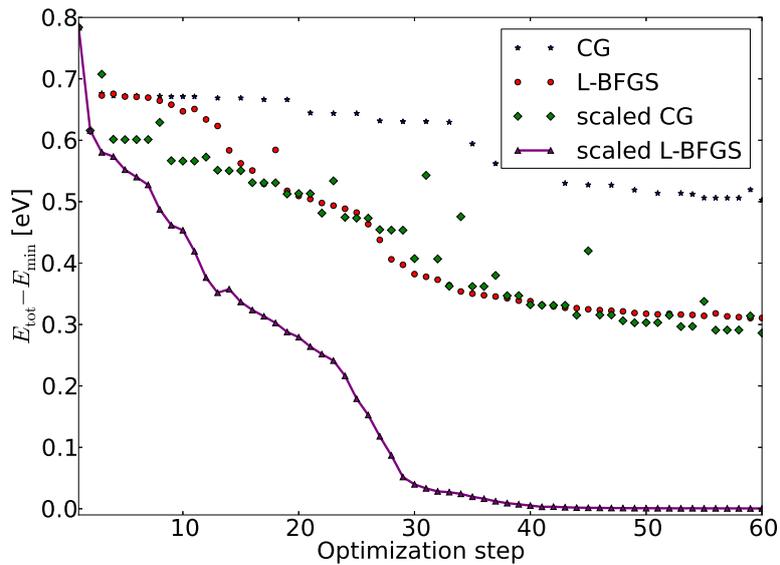}}
   \caption{Total energy over number of FOD optimization steps using different algorithms (CG, scaled CG, L-BFGS, scaled L-BFGS) for (a) Kr and (b) Cr  atom.}
   \label{fod_opt_Kr_Cr}
\end{figure}

The total energy versus the optimization step for each method is shown in Fig. \ref{fod_opt_Kr_Cr}. It is clear that scaling improves both the CG and L-BFGS algorithms.  The scaled L-BFGS performs the best of the methods tested. For both Kr and Cr, the minimum energy is reached after approximately 40 steps.  None of the other methods converges to the lowest energy within the 60 steps shown. The unscaled CG method has the slowest convergence for both Kr and Cr.  The scaled CG and unscaled L-BFGS show similar performance after 60 steps, although the scaled CG leads to a faster initial drop in energy.
\newline

To implement the scaled L-BFGS method for molecules, we simply use the $h_D$ value for the appropriate shell of the nearest atom to the FOD position.  This is clear for the core orbitals, since the core FODs have positions quite close to the nuclear position.  For valence states and particularly bonding states, the FODs can lie between atoms, potentially complicating the choice of scaling parameters; however, we find that simply using the $h_D$ value for the valence shell of the nearest atom is effective.  In tests for molecules, we obtain speedups for the scaled L-BFGS method similar to those shown in Fig.  \ref{fod_opt_Kr_Cr}.

\section{Results of applying FLO-SIC}

\subsection{Atomic orbital energies}
\label{atomic_energies}
\begin{figure}[h]
    \centering
    \includegraphics[scale=0.25]{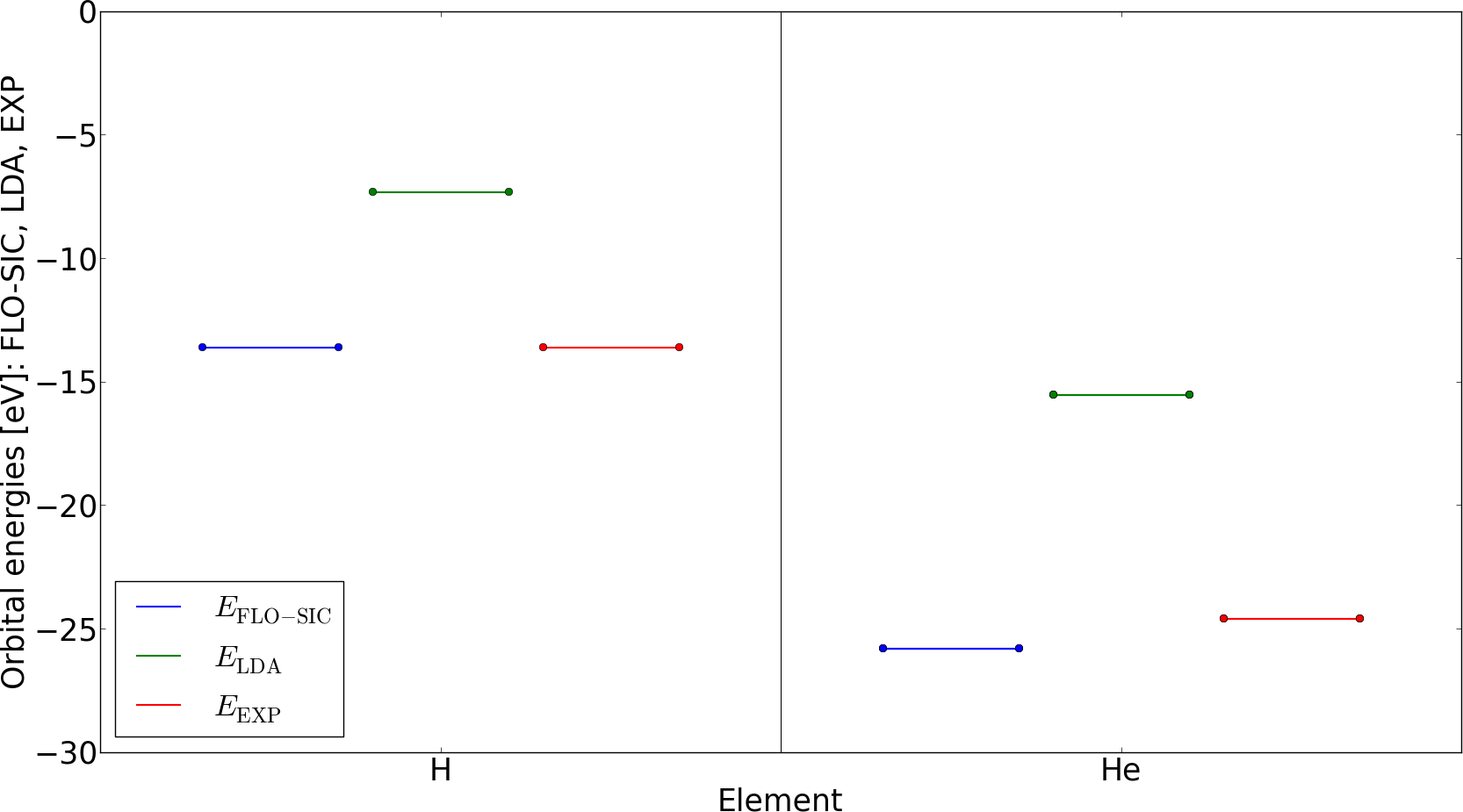}
    \caption{Atomic orbital energies for the 1s orbitals of H and He. The FLO-SIC values for H are exact while for He they are in much better agreement to experiment
    than the parent LDA functional.}
    \label{atoms_H_He}
\end{figure}

Because the potential an electron "sees" in a DFT calculation includes an interaction with itself, at a large distance $r$ from a neutral atom the potential goes exponentially to zero, 
rather than exhibiting the physically correct $-1/r$ behavior.  This leads to several problems in the description of atoms and other finite systems.  In this section we focus on orbital energies, which are too high in DFT calculations, making them poor approximations of electron removal energies. For example, the 
1s orbital in the H atom should be at $-13.6$~eV. As seen in Fig.~\ref{atoms_H_He}, the actual orbital energy in a spin-polarized DFT-LDA calculation is $-7.3$~eV. On the other hand, the FLO-SIC treatment of H (again spin-polarized), which contains only one electron, is exact and the 1s orbital energy in FLO-SIC-LDA equals the experimental removal energy. FLO-SIC is not exact for He, but the He 1s energy level is clearly brought into much better agreement with experiment in FLO-SIC-LDA than in the uncorrected theory.

\begin{figure}[h]
    \centering
    \includegraphics[scale=0.30]{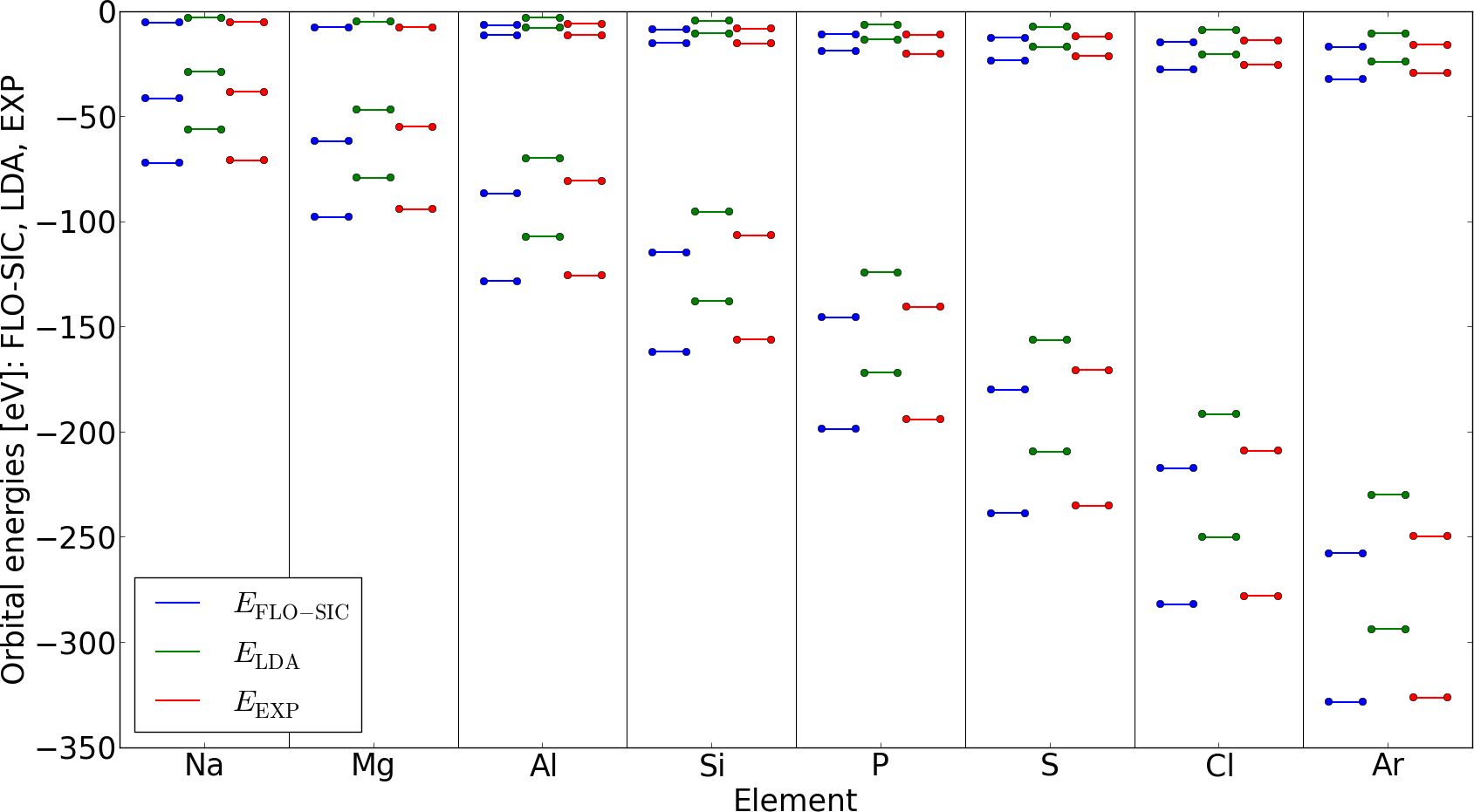}
    \caption{Atomic orbital energies for the 2s2p and 3s3p shells from LDA, FLO-SIC-LDA and experiment for Na-Ar. The FLO-SIC values are in much better agreement with experiment
    than those of the parent LDA functional.}
    \label{atoms_Na_Ar}
\end{figure}

It is interesting to look at other elements and at different energy levels (valence, semi-core, core) to determine
whether all energy levels are corrected in a FLO-SIC calculation. This is done in Fig.~\ref{atoms_Na_Ar}, where the orbital energies of the 2s2p as well as 3s3p shells for the atoms 
from Na-Ar are shown. Even these deeper energy levels are strongly corrected towards the experimental removal energies when comparing DFT-LDA and FLO-SIC-LDA. (All calculations involving open shell atoms are done spin-polarized. The splitting of the majority and minority spin levels is not visible on the scale of Fig.~\ref{atoms_Na_Ar}, though all levels are plotted.)

Taking all the orbital levels for all elements from H to Zn and comparing them to experiment, we find the errors given in table~\ref{atomic_orb_errors}. It can be seen that 
not only the value of the highest occupied orbital is corrected~\cite{atomicFODs2017,SIC-LEvsFLOSIC}, but all other orbital levels are as well, down to the deepest core levels. Accordingly, FLO-SIC restores to all orbital energies the physical significance of experimental removal energies.  
\begin{table}[h]
    \centering
    \caption{Mean absolute error (MAE) and mean absolute percentage error (MAPE) of the atomic orbital energies for H-Zn in comparison to 
    experimental values. The notation 'all' means that all orbital energies were taken into account (from valence to core) while 'HOO' refers to the highest occupied 
    orbital and '1s' to the lowest occupied orbital. The agreement of FLO-SIC with experimental data is significantly better than that of the parent LDA functional and close to Hartree-Fock~\cite{HF_levels}.}
    \begin{tabular}{|l|l|r|r|r|}
    \hline
         Orbitals                       & Error         & \multicolumn{1}{c|}{LDA}      & \multicolumn{1}{c|}{FLO-SIC-LDA}  & \multicolumn{1}{c|}{Hartree-Fock}     \\\hline
         \multirow{2}{*}{all}           & MAE [eV]      & 34.3                          & 7.2                               & 9.1                                   \\
                                        & MAPE [\%]     & 32.8                          & 5.8                               & 6.5                                   \\\hline
         \multirow{2}{*}{HOO}           & MAE [eV]      & 4.2                           & 0.7                               & 0.7                                   \\
                                        & MAPE [\%]     & 70.6                          & 6.2                               & 9.1                                   \\\hline
         \multirow{2}{*}{1s}            & MAE [eV]      & 104.9                         & 19.8                              & 18.9                                  \\
                                        & MAPE [\%]     & 9.9                           & 1.9                               & 1.8                                   \\\hline
    \end{tabular}
    \label{atomic_orb_errors}
\end{table}

\subsection{Dissociation Curves}
\begin{figure}[h]
    \centering
    \includegraphics[scale=0.25]{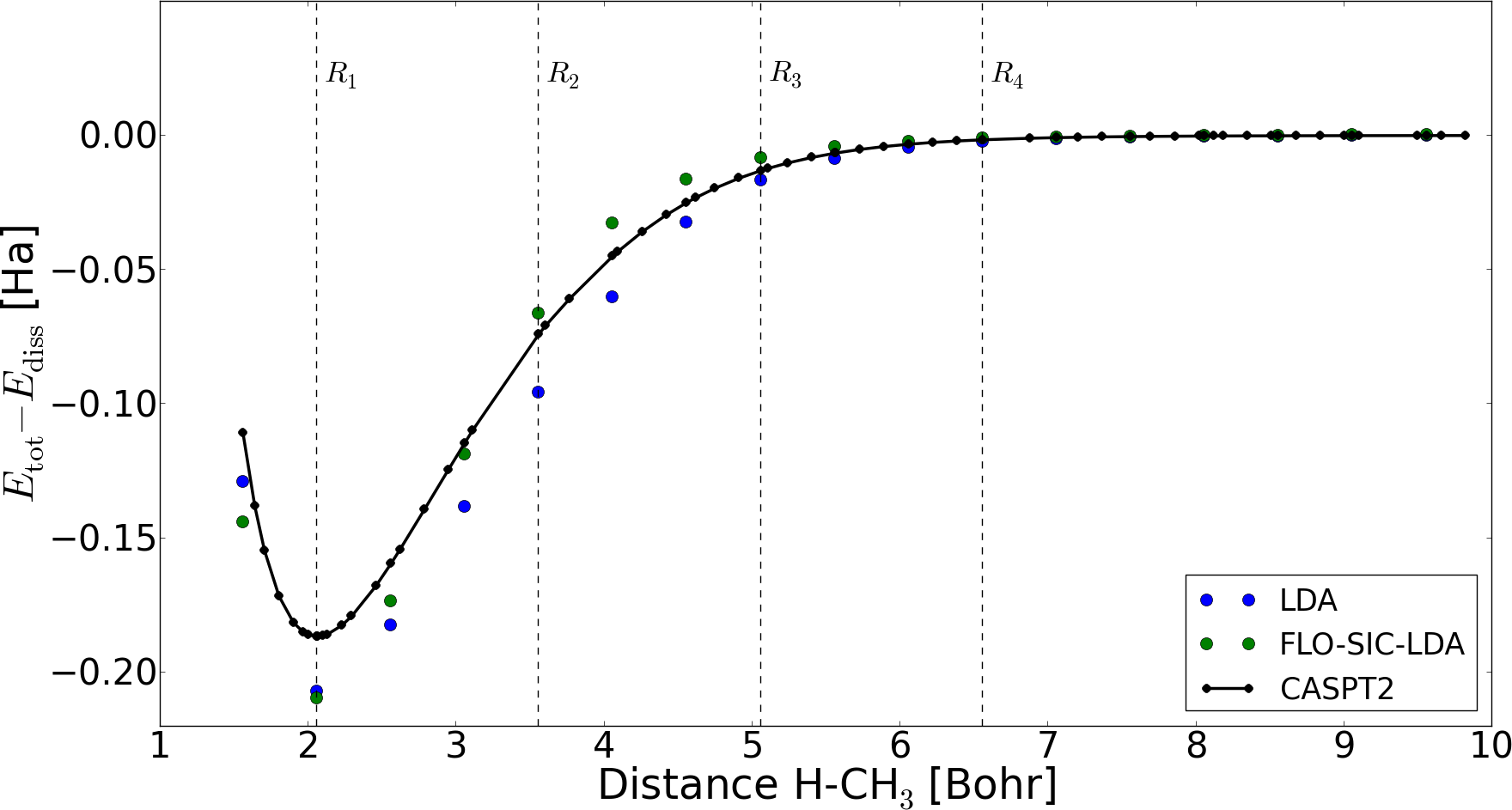}
    \caption{Energy with respect to the dissociated limit $E_{\text{diss}}$ for moving a single H away from CH$_{4}$ calculated with LDA, FLO-SIC-LDA and CASPT2. The 
    distances used in the error calculation are indicated by dashed lines.}
    \label{energy_CH4}
\end{figure}

To investigate the performance of FLO-SIC in stretched bond situations, we computed dissociation curves by systematically removing a single H from LiH, BeH${_2}$, BH$_{3}$, CH$_{4}$, NH$_{3}$, H$_{2}$O, and HF, while holding the remaining atoms fixed.  An example of such a curve for the CH$_{4}$ molecule is shown in Fig.~\ref{energy_CH4}, where a comparison between LDA, FLO-SIC-LDA and reference CASPT2 calculations is shown. The LDA and FLO-SIC-LDA calculations were done without spin polarization.  Energies are plotted with respect to the separated limit of an isolated H atom and an isolated molecular fragment.  

\begin{figure}[h]
    \centering
    \subfigure[Mean error]{\includegraphics[scale=0.20]{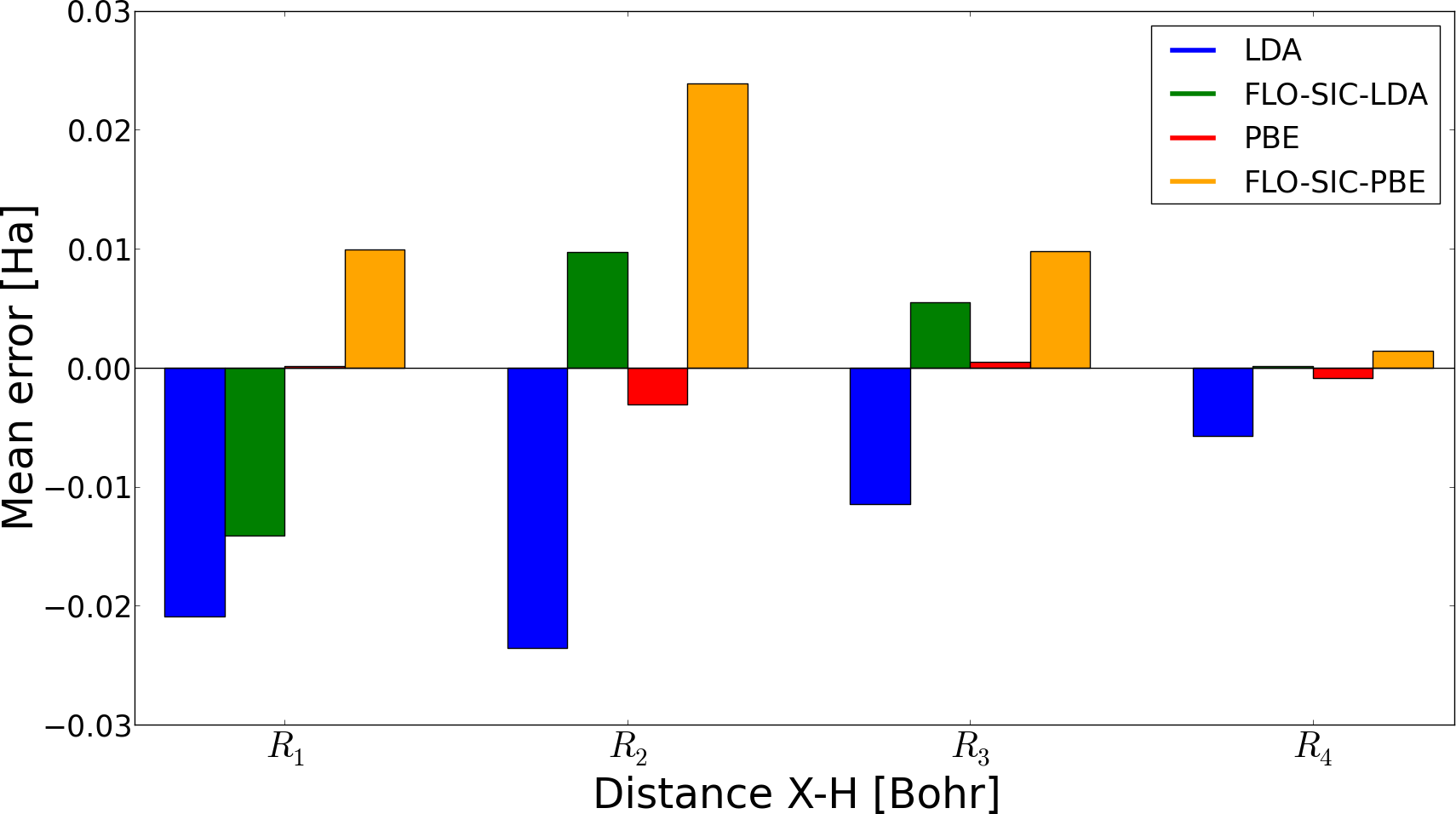}}
    \subfigure[Mean absolute error]{\includegraphics[scale=0.20]{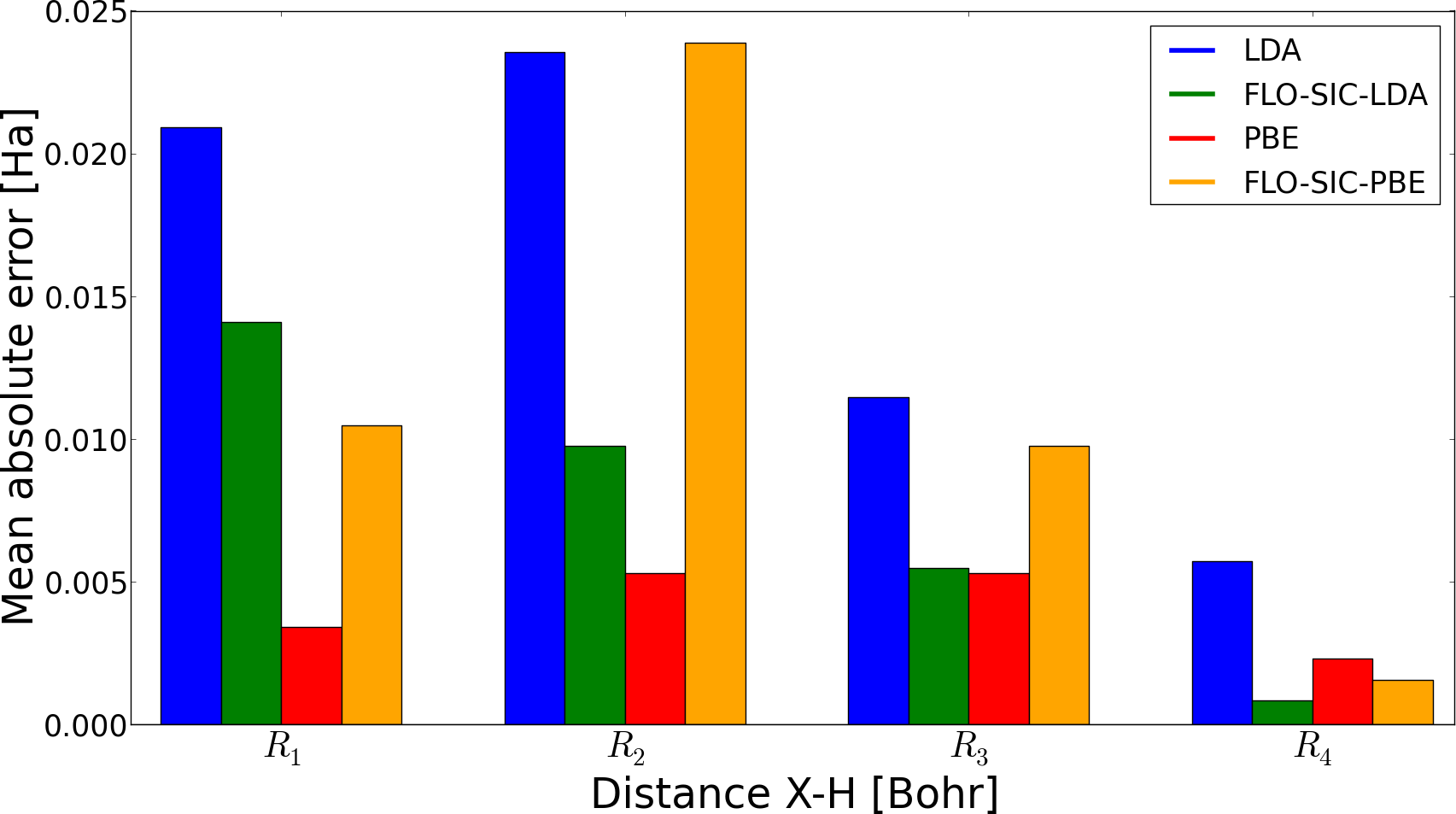}}
    \caption{Errors of LDA, FLO-SIC-LDA, PBE and FLO-SIC-PBE respect to the CASPT2 reference values at four different points along the X-H dissociation curve (see figure~\ref{energy_CH4}) for LiH, BeH$_{2}$, BH$_{3}$, CH$_{4}$, NH$_{3}$, H$_{2}$O, and HF.  See the text for more details.}
    \label{errors_CH4}
\end{figure}

For a more quantitative analysis of the results for all seven molecules, the error (calculation$-$reference) at four equidistant  points along the energy curves is determined for LDA, FLO-SIC-LDA, PBE, and FLO-SIC-PBE. The first of these points is the equilibrium X-H distance where the energy has the value $E = E_\text{b}$, the binding energy of the bond.  The second is the separation at which $E = E_{\text{b}}/2$.  The third and fourth points are then determined by making two additional steps of the same length outward from the second point.  The mean error and the mean absolute error with respect to the reference values for all seven molecules is given in Fig.~\ref{errors_CH4}. 
It can be seen that FLO-SIC-LDA always improves the LDA energy, while FLO-SIC-PBE tends to over-correct the already good description of PBE. For the largest separations, 
FLO-SIC reduces the mean absolute error in both LDA and PBE, for LDA dramatically.

\subsection{Magnetic Exchange Couplings}


As shown in Section~\ref{atomic_energies}, SIE strongly affects orbital energy levels and hence it is also expected to impact molecular properties.  
One property of particular interest for possible nanotechnology applications are magnetic exchange couplings ($J$) of transition metal systems.  The couplings 
gauge the nature and strength of spin interaction among metal centers~\cite{J_strength, J_meaning}. 
The effect of SIE on $J$ has been analyzed in the past for transition metal complexes empirically using several widely available hybrid functionals with different amount
of exact Hartree-Fock exchange~\cite{nonprojected2}. Here we show the effect of explicitly removing SIE on the magnetic
exchange couplings of the transition metal complex [Fe$_2$OCl$_6$]$^{2-}$ using the FLO-SIC method. 
 \newline

The computational burden associated with optimizing hundreds of FODs in a large system can be  reduced by using
effective core potentials (ECPs), where only the valence electrons (and hence valence FODs) are 
treated explicitly. This simple strategy enables 
calculations on large systems by substantially reducing the computational time. The use of ECPs has been shown to work well for a number of organic radicals as well as for 
another transition metal complex ([Cu$_{2}$Cl$_{6}]^{2-}$)~\cite{J_SIC}. 
Here, we  used the Stuttgart effective core potentials \cite{Feller-ECP-basis1} for Fe and Cl  atoms with the corresponding uncontracted basis sets, whereas the O atom was 
treated in an all-electron manner. We calculate the magnetic exchange couplings ($J$) using the spin-projected approach of Noodleman~\cite{Noodleman},
\begin{equation}
\label{JAB}
J_\text{{}} = \frac{E_{\text{BS}}-E_{\text{HS}}}{2  S_{A}  S_{B}},
\end{equation}

where $E_{\text{HS}}$ and $E_{\text{BS}}$ are the energies  
for the high spin state (HS) and a broken-symmetry spin state (BS), respectively, and $S_{A}$ and $S_{B}$ are
the nominal spins on the two centers.

\begin{table}[h!]
\begin{center}
\caption{Magnetic exchange couplings (cm$^{-1} $) for [Fe$_2$OCl$_6$]$^{2-}$ using different methods.}
\begin{tabular}{ lr } 
\hline
\hline
Methods         & $J_\text{}$      \\\hline
LDA            &  $-$495            \\
FLO-SIC          &  $-$97             \\ 
PBEh            &  $-$377            \\ 
B3LYP           &  $-$443            \\ 
M062X           &  $-$284            \\ 
$\omega$B97XD   &  $-$384            \\ 
Experiment~\cite{Fe2OCl6-exp}      &  $-$112            \\\hline\hline
\end{tabular}
\label{table:J}
\end{center}
\end{table}

In table~\ref{table:J}, we present the $J$ values obtained from plain LDA and FLO-SIC-LDA (dubbed FLO-SIC in the table),
along with the ones from multiple hybrid functionals.  The experimental value is listed as a reference. The $J$ value with plain LDA
is overestimated by a large amount when compared to the reference value. However, with FLO-SIC it is
significantly corrected and the resulting $J$ value is close to the one obtained from experiment. 
All hybrid functionals considered in this work correct the exchange coupling toward the experimental reference similar to FLO-SIC and lie between the LDA and experimental values. 
The hybrid functionals partially correct for self-interaction  by including a  fraction of exact Hartree-Fock exchange in the exchange-correlation energy.

\subsection{Charge transfer and self-interaction}
SIE also impairs the description of charge transfer in DFT calculations.  For example, it has been shown that heteronuclear dimers dissociate to fractionally charged atoms in DFT, due to unphysical positions of the atomic energy levels caused by self-interaction~\cite{adrienn}. A related, but more subtle effect is the description of the charge distribution in polar molecules at equilibrium bond lengths.  For molecules near equilibrium, PBE gives a good description of properties like binding energies, yet properties like dipole moments do not agree well with experimental values.  This is a result of electron self-interaction that makes the anions relatively less stable in DFT calculations, resulting in less charge transfer and smaller dipole moments.  This can be seen in table~\ref{table:dipoles} where results from PBE and FLO-SIC-PBE are compared with reference CCSD(T) values~\cite{head-gordon}.  Experimental geometries were used for all calculations.  PBE results in dipole moments that are systematically too small, while  FLO-SIC-PBE produces values in much closer agreement with the reference values.  For the six molecules shown in the table, the average PBE result underestimates the reference values by 5.2\%, while the FLO-SIC-PBE values have an average error of only 1.8\%, an improvement of roughly 3 times over PBE. 
\newline

We also show calculated Mulliken charges for the anions in table~\ref{table:dipoles}.  The increase in the dipole moments in FLO-SIC-PBE is clearly due to an increase in the charge transferred to the anion.  For the six molecules shown, the charge on the anion is greater on average by 0.07 electrons.  

\begin{table}[h!]
\begin{center}
\caption{Dipole moments (in Debye) and Mulliken charge ($Q$, in electrons) on the anion computed for ionic molecules in PBE and FLO-SIC-PBE.  
Reference values for the dipole moments were obtained using the CCSD(T) method~\cite{head-gordon}.}
\begin{tabular}{ l c c c | c c} 
\hline
\hline
& \multicolumn{3}{ c }{Dipole (Debye)} & \multicolumn{2}{ |c }{$Q$ (e)}  \\\hline
     & PBE   & FLO-SIC-PBE & CCSD(T) & PBE & FLO-SIC-PBE \\\hline
LiH	 & 5.570 & 5.943 & 5.829 & 0.477 & 0.585 \\
LiF  & 6.129 & 6.347 & 6.288 & 0.833 & 0.889 \\
LiCl & 6.890 & 7.281 & 7.096 & 0.726 & 0.773 \\
NaH  & 5.751 & 6.689 & 6.397 & 0.413 & 0.497 \\
NaF  & 7.698 & 8.191 & 8.134 & 0.814 & 0.896 \\
NaCl & 8.475 & 8.984 & 9.007 & 0.721 & 0.757

\\\hline\hline
\end{tabular}
\label{table:dipoles}
\end{center}
\end{table}

\section{Summary and future directions}
The applications of the FLO-SIC method presented above illustrate the success of FLO-SIC and the PZ-SIC formalism in removing the effects of SIE from commonly used exchange-correlation functionals in situations where these errors are important.  Similar improvements over uncorrected functionals have been found in other benchmarking applications carried out by our group~\cite{kamal, PhysRevA.100.012505} and others~\cite{G2}.  Challenges remain to making FLO-SIC more efficient for practical calculations.  We are currently working on creating improved starting points for FOD positions, based on optimized results for the free atoms and accumulating experience gained in describing molecules.  Good starting points can greatly reduce the number of optimization steps needed to find energy-minimizing FLOs.  Further improvements to the optimization schemes are also possible, with the goal of accelerating convergence still further.  We are also pursuing algorithmic improvements to make the solution of the DFT-SIC equations more efficient.  Also needed is the implementation of periodic boundary conditions, to make FLO-SIC calculations possible for crystaline materials.  A more fundamental goal is to improve the performance of FLO-SIC for problems where the underlying DFT functional already works well.  As illustrated by the data for $R_1$ in Fig.~\ref{errors_CH4}, FLO-SIC-PBE yields worse binding energies than PBE for molecules near their equilibrium geometries.  More work is needed to determine how to best pair SIC with functionals such as PBE and the promising new SCAN functional~\cite{SCAN} in order to remove SIE when it is prominent, without diminishing their performance when it is not.  This, combined with more efficient computational tools, can make self-interaction-free DFT calculations a reality.  
\newline

\section{Acknowledgments}

We gratefully acknowledge the financial support of the U.S. Department of Energy, Office of Science, Office of Basic Energy Sciences as part of the Computational Chemical Sciences Program under award \#DE-SC0018331.
\newcommand{\newblock}{}
\bibliographystyle{iopart-num}
\bibliography{ref}

\end{document}